\documentclass[iop, apj]{emulateapj}
\usepackage{graphicx}

\begin{document}


\def\ba{\begin{eqnarray}}
\def\ea{\end{eqnarray}}
\def\etal{et al.\ \rm}
\def\tpar{\tau_\parallel}
\def\accom{\langle\alpha\rangle}

\newcommand\tna{\,\tablenotemark{a}}
\newcommand{\superscript}[1]{\ensuremath{^{\textrm{\scriptsize{#1}}}}}
\newcommand{\subscript}[1]{\ensuremath{_{\textrm{\scriptsize{#1}}}}}

\title{Inner edges of compact debris disks around metal-rich white dwarfs.}

\author{Roman R. Rafikov\altaffilmark{1,2}}
\author{Jos\'{e} A. Garmilla\altaffilmark{1}}
\altaffiltext{1}{Department of Astrophysical Sciences, 
Princeton University, Ivy Lane, Princeton, NJ 08540; 
rrr,garmilla@astro.princeton.edu}
\altaffiltext{2}{Sloan Fellow}


\begin{abstract}
A number of metal-rich white dwarfs (WDs) are known to host
compact, dense particle disks, which are thought to be 
responsible for metal pollution of these stars. In many such
systems the inner radii of disks inferred from their spectra 
are so close to the WD that particles directly exposed to
starlight must be heated above $1500$ K and are expected  
to be unstable against sublimation. To reconcile this expectation with 
observations we explore particle sublimation in H-poor 
debris disks around WDs. We show that because of the high 
metal vapor pressure the characteristic sublimation 
temperature in these disks is $300-400$ K higher than in their
protoplanetary analogues, allowing particles to survive
at higher temperatures. We then look at the structure of the 
inner edges of debris disks and show that they should generically 
feature superheated inner rims directly exposed to starlight with 
temperatures reaching $2500-3500$ K. Particles migrating 
through the rim towards the WD (and rapidly sublimating) 
shield the disk behind them from strong stellar heating, making 
the survival of solids possible close to the WD. Our model 
agrees well with observations of WD+disk systems provided 
that disk particles are composed of Si-rich material such 
as olivine, and have sizes in the range $\sim 0.03-30$ cm.
\end{abstract}



\section{Introduction.}  
\label{sect:intro}


About two dozen (at the moment of writing) white dwarfs (WDs)
are known to exhibit near-IR excesses in their spectra (e.g.
\citealt{zuckerman_1987}; \citealt{kilic_2005}, 2006; \citealt{jura_2009},
\citealt{farihi_2009}). This is usually interpreted (\citealt{graham_1990};
\citealt{jura_jan2003}, 2006; \citealt{farihi_2009}) as evidence for the 
existence of nearby solid debris reprocessing stellar 
radiation in the IR. 
Detailed spectral modeling of excesses generally supports the idea that 
debris particles are arranged in a disk-like configuration, which
is optically thick but geometrically very thin\footnote{Some 
exceptions are also known such as HD233517, the spectrum of which 
is better fitted by invoking a flared disk (\citealt{jura_jan2003}), or 
GD56, which is better fitted by a warped disk (\citealt{jura_2009}).}, 
thus having properties very similar
to the rings of Saturn (\citealt{cuzzi_2010}). 
These disks are relatively compact, $\lesssim 1$ R$_\odot$, 
excluding the possibility of a primordial 
origin because of the long cooling ages of the WDs around
which they are found, $\sim 0.1-1$ Gyr. It was proposed by
\citet{jura_feb2003}, following an earlier suggestion by \citet{alcock_1986}, 
that these disks are produced by tidal disruption of 
asteroid-like bodies launched on low-periastron orbits by 
distant massive planets \citep{debes_2002}, 
which have survived the asymptotic giant branch (AGB) phase of the 
evolution of the central star.

All WDs possessing compact debris disks exhibit atmospheres which are 
polluted (sometimes heavily) with metals, putting these WDs into 
the DAZ and DBZ classes \citep{farihi_mar2011}. This observation strongly 
suggests that
many (if not all) metal-rich WDs are polluted by accretion of high-Z
elements from the compact debris disks around them. In cases where
near-IR observations do not reveal the presence of a conspicuous disk
of solids, the disk may simply be too tenuous to reprocess enough 
stellar radiation to make itself visible. Alternatively, the
disk of solids may have dispersed some time ago but the metals
can still be present in the WD atmosphere because of their long 
settling time \citep{metzger_2012}.

In this scenario of WD metal pollution (which is certainly 
valid for systems with known 
disks) the issue of metal transfer onto the WD surface must be 
addressed, as the disk of solids cannot extend inward all the way 
to the stellar surface at $R_\star$. Because of the high effective 
temperature of these WDs, $T_\star\sim (7-20)\times 10^3$ K, solids
must sublimate at some inner radius $R_{in}$ producing metal 
gas, which is subsequently accreted onto the WD via a conventional 
accretion disk. The existence of inner cavities in disks of 
solid debris follows directly from the shape of their spectral
energy distributions (SEDs), which generally show a lack of emission 
corresponding to temperatures in excess of $\sim 2000$ K. 

It has been shown by \citet{rafikov_may2011} and
\citet{bochkarev_2011}
that Poynting-Robertson (PR)drag on the disk of 
particles naturally drives accretion of solids at rates 
$\dot M_Z \sim 10^7-10^8$ g s$^{-1}$. Their sublimation feeds metal
gas accretion onto the WD surface at the same rate. Even higher 
$\dot M_Z$ can be achieved if the disk of solids can couple 
via aerodynamic drag to the surrounding gaseous disk 
(\citealt{gansicke_2006}, 2007; \citealt{brinkworth_2009};
\citealt{melis_2010}), which naturally forms via
sublimation of particles at $R_{in}$ (\citealt{rafikov_sep2011}; 
\citealt{metzger_2012}).


\subsection{Inner rim puzzle}
\label{sect:rimpuzzle}

Existing debris disk models used to fit SEDs try to 
link the radius of the inner edge of the disk $R_{in}$ to
a certain value of the ``sublimation temperature'' $T_s$, 
which depends on physical properties of the constituent 
particles. In these models particles sublimate at radii
where their temperature $T$ exceeds the sublimation 
temperature $T_s$, and the inner radius $R_{in}$ 
corresponds to $T(R_{in})=T_s$. Then the determination of 
$R_{in}$ hinges upon the proper choice of $T_s$ and the 
knowledge of a relation between $T$ and $r$. 

The characteristic value of $T_s$ usually assumed for 
disks around WDs is $1300-1500$ K. This is a typical 
sublimation temperature for the Si-rich solids, such as olivine 
or calcium-aluminum inclusions (CAIs) based on \citet{lodders_2003}
who calculated condensation temperatures 
of different species in the proto-Solar disk 
assuming Solar abundances of elements. There is good 
evidence that Si-rich material indeed represents a 
significant fraction of mass in the debris disks around 
WDs, both from the 
detections of the Si feature at 10 $\mu$m in spectra of 
such disks \citep{jura_2009} and the atmospheric 
compositions of their host  
WDs (\citealt{zuckerman_2007}; \citealt{klein_2010}, 2011;
\citealt{jura_may2012}). However, as we show in 
\S \ref{sect:sublimation} this estimate 
of $T_s$ needs to be seriously revised for 
the typical conditions in circum-WD debris disks. 
 
There is certain ambiguity regarding the equilibrium 
temperature of the disk particles. In Appendix \ref{sect:thermal}
we consider their thermal balance by looking 
at different heating and cooling processes, and find
stellar heating and radiative cooling of particles to 
dominate the balance. In this case particles directly 
exposed to starlight or located in the optically thin 
parts of the disk, are heated to a temperature
\ba
T_{\rm thin}(r)\approx T_\star\left(\frac{R_\star}{2r}\right)^{1/2}. 
\label{eq:Tofr}
\ea
In particular, this estimate is appropriate for 
particles at the inner edge of the optically thick 
disk because these are directly illuminated 
by the star. 
However, behind the narrow rim of directly exposed particles the 
disk is illuminated by starlight only at its surface at a grazing 
incidence angle $\zeta\approx (4/3\pi)R_\star/r$ \citep{friedjung_1985}
for a geometrically thin (i.e. flat) disk. This is because of 
the shielding that is provided by the rim particles against 
direct starlight for the disk just outside the rim, see 
Figure \ref{fig:opt_thick} for illustration. The equilibrium 
temperature of particles in the optically thick parts of 
the disk is given by \citep{chiang_1997}
\ba
T_{\rm thick}(r)=T_\star\left(\frac{2}{3\pi}\right)^{1/4}
\left(\frac{R_\star}{r}\right)^{3/4}.
\label{eq:Tflat}
\ea  
This expression is valid in the shielded parts of the disk where 
the near-IR emission is produced. 

The assumption that particles sublimate at a single temperature 
$T_s$ implies that directly exposed rim particles cannot 
be hotter than $T_s$ (\citealt{rafikov_may2011}, b). However, using 
equation (\ref{eq:Tofr}) 
to estimate $T(r)$ at the inner edge of the disk and taking 
$T_s\approx 1500$ K one finds that the near-IR contribution 
to the SED produced by the disk is too weak to account for
observations. Indeed, combining equations (\ref{eq:Tofr}) 
and (\ref{eq:Tflat}) one finds
\ba
T_{\rm thick}=\left(\frac{16}{3\pi}\right)^{1/4}
T_{\rm thin}\left(\frac{T_{\rm thin}}{T_\star}\right)^{1/2}.
\label{eq:Trat}
\ea
Thus, when $T_{\rm thin}$ is close to the sublimation
temperature the temperature $T_{\rm thick}$ in the 
shielded part of the disk must be substantially lower 
than $T_s$ (since necessarily 
$T_{\rm thin}<T_\star$). For example, taking $T_\star=10^4$ K and 
$T_{\rm thin}(R_{in})=T_s=1500$ K one finds 
$T_{\rm thick}(R_{in})\approx 660$ K. As a result, the SED 
of such a disk is going to be very deficient of the near-IR 
flux corresponding to emission at temperatures 
$\gtrsim 1300-1500$ K. However, disk
SEDs typically exhibit considerable emission by material 
heated in excess of $1000$ K, see Table \ref{tbl:systems}. 
This is hard to reconcile with only the inner rim of the 
disk being heated to $T_s$. 

\citet{jura_2012} suggested that this problem can be 
resolved if the inner edge of the disk is set by sublimation 
occurring in the {\it optically thick} part of the disk illuminated 
by the star at grazing incidence, rather than in the thin
inner rim of directly exposed particles. This is equivalent to
determining the value of $R_{in}$ by using the expression 
(\ref{eq:Tflat}) instead of (\ref{eq:Tofr}) in equation 
$T(R_{in})=T_s$.
If that were true, however, then the temperature at the inner 
rim must be higher than $T_s$, see equation (\ref{eq:Trat}), 
and rim particles would be 
sublimating, exposing the particles behind them to direct 
starlight. As a result, the inner rim would recede to larger 
distance from the WD until it reaches the radius where 
$T_{\rm thin}=T_s$, so we go back to the previously 
considered situation with its intrinsic problems. 
Only this configuration is going to be in stable phase equilibrium 
as long as sublimation is idealized as a step-like
process, i.e. that particles turn into gas
as soon as they reach $T_s$. Such an equilibrium 
was assumed in Rafikov (2011a, b) to determine $R_{in}$. 

This set of conflicting arguments suggests that our
understanding of the location and structure of the 
inner rim is in some ways incomplete. The goal of 
the present work is to fill these gaps and to provide 
a more detailed picture of the sublimation of solids 
at the inner edge of the disk by focussing on two effects. 
First, in \S \ref{sect:sublimation} we show that 
sublimation in hydrogen-poor debris disks around WDs is different 
from sublimation in the proto-Solar disk resulting 
in $T_s$ being higher than 1500 K. Second, we show in 
\S \ref{sect:thick} that particle sublimation at 
the inner rim of an optically thick disk 
is a dynamic process, which makes it possible for the rim 
particles to reach temperatures in excess of $T_s$ before 
sublimating. In \S \ref{sect:opt_thin} we look at sublimation
in optically thin disks. We apply our theory to observed 
disk-hosting WDs in \S \ref{sect:apps}, and discuss our findings
in \S \ref{sect:disc}. A summary of our main results can be found
in \S \ref{sect:sum}.


\section{Specifics of sublimation in the circum-WD disks.}
\label{sect:sublimation}

Solid particles of certain composition surrounded by vapor 
with the same elemental abundance grow by 
condensation of molecules or atoms arriving at their surfaces
from the gas phase and lose mass due to sublimation.
For a particle of mass $m_p$ and surface area $S_p$ surrounded by 
vapor at pressure $P_{\rm vap}$ one can write the 
following mass evolution equation \citep{guhathakurta_1989}:
\ba
\frac{dm_p}{dt}= S_p\left[\accom P_{\rm vap}\left(\frac{\mu}{2\pi k_B T}
\right)^{1/2}-\dot m(T)\right]. 
\label{eq:mass_balance}
\ea
where $\mu$ is the mean molecular weight of the particle material.
Here the first term in brackets describes condensation ($\accom$ 
is the accommodation coefficient --- sticking probability of gas 
particles impacting the solid surface), while the second term 
characterizes sublimation from the particle surface ($\dot m$
is the mass loss rate per unit surface area due to sublimation). 

When the vapor pressure becomes equal to the saturated vapor 
pressure $P_{\rm vap}^{\rm sat}(T)$ at a given temperature $T$, the 
equilibrium between the loss and gain processes is established 
and $dm_p/dt=0$. This allows us to express
\ba
\dot m(T)=\accom
P_{\rm vap}^{\rm sat}(T)\left(\frac{\mu}{2\pi k_B T}\right)^{1/2}.
\label{eq:mdotPvap}
\ea

The concept of sublimation temperature $T_s$ implies the process
of phase transition from solid to gas to occur in a 
step-like fashion. At $T=T_s$ the vapor saturates 
and an infinitesimal increase of temperature leads to 
slow (quasi-static) conversion of solid into gas. 
Then one can again assume $dm_p/dt\to 0$ and the 
right-hand side of equation (\ref{eq:mass_balance}) then provides
us with an implicit relation for $T_s$ as a function of the vapor
pressure $P_{\rm vap}$, as long as the dependence $\dot m(T)$
is known. 

Considerations based on the Clausius-Clapeyron relation suggest that 
\ba
P_{\rm vap}^{\rm sat}(T)\propto T^\beta e^{-T_0/T},
\label{eq:C-K}
\ea
where the constants $\beta$ and $T_0$ are specific to a 
particular particle composition. Since the strongest dependence 
of $P_{\rm vap}^{\rm sat}$ on $T$ occurs through the exponential 
factor, the power-law dependence on $T$ in equation 
(\ref{eq:mdotPvap}) can be absorbed into the (approximately) 
constant pre-factor, so that $\dot m(T)$ is approximated as 
\ba
\dot m(T)=\accom K_0 e^{-T_0/T},
\label{eq:mdot}
\ea
where $K_0$ is a constant.
This is the form of $\dot m(T)$ that we adopt in this work.

In the following we will consider a number of different materials
which can represent the composition of disk particles. We summarize 
the parameters $K_0$ and $T_0$ for different species considered 
in this work in Table \ref{tbl:materials}, and provide details 
of their calculation in Appendix \ref{app:sublimation}.

Setting the left hand side of equation (\ref{eq:mass_balance}) to zero
and using $\dot m$ in the form (\ref{eq:mdot}) we obtain the 
(implicit) dependence of $T_s$ on the vapor pressure:
\ba
T_s(P_{\rm vap}) &=& T_0\left(\ln\Lambda_s\right)^{-1},
\label{eq:T_s_P}\\
\Lambda_s &=& \frac{K_0}{P_{\rm vap}}
\left(\frac{2\pi k T_s}{\mu}\right)^{1/2},
\label{eq:Lambda_s}
\ea
with $\Lambda_s\gg 1$. 
According to this expression $T_s$ is higher for larger 
$P_{\rm vap}$, even though the dependence is rather weak
(logarithmic).  

The vapor pressure in the gas around the rim can be easily 
estimated if the disks consist of particles with identical 
composition. In general this does not have to be true but we 
still adopt this assumption for simplicity. Then metals 
detected in the WD atmosphere come from accretion 
of this material in the gas phase, and the measurement of the 
corresponding mass accretion rate
\ba
\dot M_Z=3\pi\nu\Sigma_g
\label{eq:Mdot}
\ea
(where $\nu=\alpha_\nu c_s^2/\Omega$ 
is the kinematic viscosity, $\alpha_\nu$ is the effective viscosity 
parameter, $\Omega$ is the Keplerian 
angular frequency, and $c_s$ and $\Sigma_g$ are the sound speed 
and the surface density of the gas) provides an
estimate of the vapor pressure:
\ba
P_{\rm vap}&\approx &\frac{\dot M_Z\Omega^2}{3\pi\alpha_\nu c_s}
\label{eq:Pvap}\\
&\approx & 
0.5~\mbox{dyne cm}^{-2}~\frac{\dot M_{Z,8}M_{\star,1}}
{\alpha_{\nu,-2}c_{s,1}}
\left(\frac{0.2R_\odot}{r}\right)^{3}.
\nonumber
\ea
Here $\dot M_{Z,8}\equiv\dot M_Z/(10^8$ g s$^{-1})$, $M_{\star,1}\equiv
M_\star/M_\odot$, $\alpha_{\nu,-2}\equiv \alpha_\nu/10^{-2}$, and 
$c_{s,1}\equiv c_s/(1$ km s$^{-1})$ is the characteristic 
value of the sound speed for Si-rich material heated to 
temperature of several $10^3$ K. According to equation 
(\ref{eq:Pvap}) $P_{\rm vap}\sim 1$ dyne 
cm$^{-2}$ for $\dot M_Z\sim 10^8$ g s$^{-1}$, which is a 
characteristic mass accretion rate of metals due to the PR 
drag \citep{rafikov_may2011}. It should also be mentioned that a number of 
metal-rich WDs with debris disks exhibit much higher values of 
$\dot M_Z$, easily reaching $10^9-10^{10}$ g s$^{-1}$. In these 
systems $P_{\rm vap}\sim 10-100$ dyne cm$^{-2}$ 
should be typical.

\begin{center}
\begin{deluxetable}{lcccc}
    \tablewidth{0pt}
\tablecaption{Sublimation properties of different materials
\label{tbl:materials}}
    \tablehead{
      \colhead{Material} &
      \colhead{$K_{0}$} &
      \colhead{$T_{0}$} &
      \colhead{$\mu$} &
      \colhead{$T_{\textrm{\scriptsize{sub}}}(1~\mbox{dyne cm}^{-2})$} \\
      \colhead{} &
      \colhead{g\superscript{-1}cm\superscript{-2}s\superscript{-1}} &
      \colhead{K} &
      \colhead{m\subscript{p}} &
      \colhead{K}
      }
    \startdata
      Olivine \hspace{5 pt} & \hspace{3pt} $1.6 \times 10^{9}$ \hspace{3pt} &
      \hspace{3pt} 68100 \hspace{3pt} & \hspace{3pt} 141 \hspace{3pt} &
      \hspace{0pt}  2100 \hspace{0pt}\\
      Graphite & $9.2 \times 10^{7}$ & 81200 & 12 
      & 2600\\
      CAI & $1.1 \times 10^{10}$ & 69400 & 274 & 2000\\
      Iron & $2.3 \times 10^{7}$ & 45400 & 56 &  1600 \\
      Al\subscript{2}O\subscript{3} & $8 \times 10^{9}$ & 80500 & 102
      & 2300 \\
      SiC & $6 \times 10^{8}$ & 73700 & 40 & 2300 \\
    \enddata
\end{deluxetable}
\end{center}

In Table \ref{tbl:materials} we show the values of $T_s$
computed from equation (\ref{eq:T_s_P}) for different materials
assuming $P_{\rm vap}=1$ dyne cm$^{-2}$. One can see that 
these values are considerably higher than the conventional 
estimate $T_s\sim 1500$ K often used in modeling debris disk
SEDs. The explanation for this puzzling difference lies in 
the fact that the canonical estimate is based on the 
work of \citet{lodders_2003} which explicitly assumes a Solar 
composition gas in equilibrium with sublimating particles to
compute $T_s$. Even though a total pressure of $100$ 
dyne cm$^{-2}$ is assumed in that work the {\it vapor} pressure
$P_{\rm vap}$
of high-Z species is going to be much lower in protoplanetary 
disks because of the low abundance of such elements compared 
to H, which contributes most to the total pressure. For 
example, in \citet{lodders_2003} iron was assumed to have an 
abundance (by number, with respect to H) of 
$3.4\times 10^{-5}$, which results in vapor pressure of 
atomic Fe of $3.4\times 10^{-3}$ dyne cm$^{-2}$ (assuming 
that molecular H has dissociated and the total
pressure in the gas is $100$ dyne cm$^{-2}$). Using equation 
(\ref{eq:T_s_P}) we then find $T_s\approx 1300$ K instead of
1600 K typical for a debris disk around a WD, if the 
latter were composed 
of pure Fe and had a total pressure (equal to the Fe vapor 
pressure) of $1$ dyne cm$^{-2}$. A similar or even larger 
difference with the canonical estimate of $T_s$ arises 
for other elements listed in Table \ref{tbl:materials}.

A deficiency of volatile components (H and He) and high 
relative abundance of metals (can easily be as high as unity) 
in the gaseous phase of the debris disks around WDs naturally 
results in high values of the (quasi-static) sublimation 
temperature $T_s$ in these systems. This goes in the direction 
of alleviating the puzzle of high temperatures of solid 
particles inferred from the SED modeling for these objects. 
However, it does not fully resolve this problem because in the
simple model of sublimation $T_s$ is still reached {\it only in 
the inner rim} of the disk, i.e. $T_s=T_{\rm thin}(r_{\rm rim})$. 
The temperature in the bulk of the disk just behind the 
rim is again much lower than  $T_s$: for
$T_\star=10^4$ K and $T_s=2100$ K as typical for olivines 
(see Table \ref{tbl:materials}) one finds using equation 
(\ref{eq:Trat}) that $T_{\rm thick}(r_{\rm rim})\approx 1100$ 
K, which is clearly not 
enough to reproduce the short-wavelength portion of the observed
SED in many WD+disk systems, see Table \ref{tbl:systems}.

In the following section we provide a complete solution to 
this puzzle by considering particle sublimation in more detail.


\section{Inner edge structure in the optically thick disk}
\label{sect:thick}

\begin{figure}
\plotone{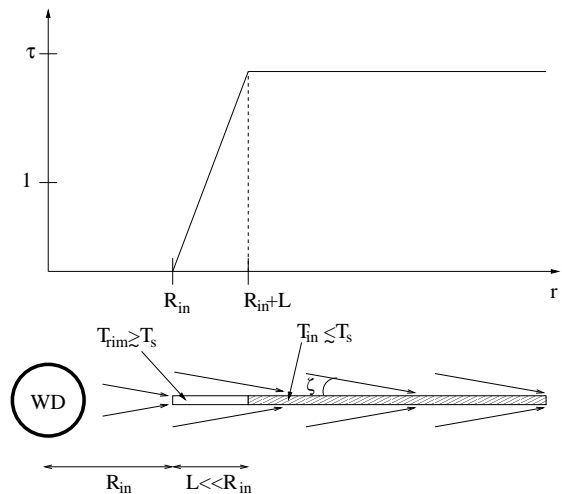}
\caption{
Schematic representation of the inner rim structure and
surface density distribution in its vicinity in the 
optically thick ($\tau\gtrsim 1$) case. The shaded part of 
the disk heated to $T_{in}\lesssim T_s$ receives 
starlight only at the surface, at 
grazing incidence angle $\zeta$. Particles in the inner, 
unshaded part are directly illuminated by the star
and are heated to $T_{\rm rim}\gtrsim T_s$. Radial width 
$L$ of this exposed rim is determined by equation 
(\ref{eq:tau=1}).
\label{fig:opt_thick}}
\end{figure}

We now present a simple physical model for the structure of
the inner rim of the disk, the inner part of which is 
{\it optically thick}, see Figure \ref{fig:opt_thick}: the vertical optical
depth of such a disk
\ba
\tau\equiv \frac{3}{4}\frac{\Sigma}{\rho a}
\label{eq:tau}
\ea
($a$ and $\rho$ are the particle radius and bulk density,
$\Sigma$ is the surface density of the disk of solids)
is larger than unity. This implies that the disk absorbs 
all incident stellar radiation because its optical depth
to starlight \citep{rafikov_may2011} 
$\tau_\parallel\equiv \tau/\zeta\gg 1$.  
According to \citet{rafikov_sep2011} and \citet{metzger_2012} such 
optically thick inner regions are natural for massive 
debris disks in which aerodynamic coupling to the 
surrounding gaseous disk is strong enough to drive 
runaway accretion of metals onto the WD. In this case 
the particle surface density at $R_{in}$ is high enough 
for the inner disk to stay optically thick.

The key ingredients of our model are 
\begin{itemize}
\item the {\it inward migration} of particles across 
the rim region,
\item the {\it shielding} from starlight provided by the 
directly exposed rim particles to particles further out,
\item the {\it dynamic} regime of particle sublimation, 
as opposed to the quasi-static situation 
explored previously. 
\end{itemize}
Our primary goal here is to determine 
the temperature inside the rim $T_{\rm rim}$ 
and the distance $R_{in}$ at which the disk gets truncated 
by sublimation. 

In the previous section, we assumed that particle 
sublimation occurs at a single temperature
$T_s$, so that particles cannot exist in 
solid form at $T>T_s$. This, 
however, is not true on time intervals shorter than 
the time it takes to completely sublimate a particle.  
Using equation (\ref{eq:mass_balance}) one can estimate 
the instantaneous {\it sublimation timescale} as the time it 
takes to completely sublimate a particle (neglecting 
condensation) at a given temperature
\ba
t_s(T)\equiv \frac{m_p}{S_p \dot m(T)}.
\label{eq:subl_time}
\ea
Sublimation should be considered as 
a {\it dynamic} (as opposed to quasi-static, like in the 
previous section) process whenever the particle temperature 
changes on a time scale $\lesssim t_s(T_s)$.

When the temperature of a solid object is close to its sublimation 
temperature $T_s(P_{\rm vap})$ (for a given vapor pressure 
$P_{\rm vap}$, which at $T_s$ should be equal to 
$P_{\rm vap}^{\rm sat}$) equations (\ref{eq:mdotPvap}), 
(\ref{eq:mass_balance}) and (\ref{eq:subl_time}) allow us to
estimate 
\ba
t_s(T_s)&\approx &\frac{a_0\rho}{3 \accom P_{\rm vap}}
\left(\frac{2\pi k_B T}{\mu}\right)^{1/2}
\label{eq:tsTs}\\
&\approx & 7~\mbox{d}~
\frac{a_{0,1} P_1\rho_1}{\accom_{0.1}\mu_{28}^{1/2}}
\left(\frac{T_s}{2000~\mbox{K}}\right)^{1/2},
\nonumber
\ea
(for a spherical particle of initial radius $a_0$)
where $a_{0,1}\equiv a_0/(1$ cm$)$, $P_1\equiv P_{\rm vap}/(1$ dyne 
cm$^{-2})$, $\rho_1\equiv (1$ g cm$^{-3})$, 
$\accom_{0.1}\equiv \accom/0.1$, and $\mu_{28}\equiv\mu/(28m_p)$,
as appropriate for Si. Because of the rapid scaling of $\dot m(T)$ 
with $T$ it is
obvious that $t_s(T)$ is a very sensitive function of $T$,
and $t_s(T)\ll t_s(T_s)$ even if $T$ is just slightly higher 
than $T_s$.

As mentioned in \S \ref{sect:rimpuzzle}, in the optically thick 
disk particles just outside the rim are shielded from direct 
starlight by the rim particles and their temperature is given by 
$T_{in}=T_{\rm thick}(R_{in})$, see equation (\ref{eq:Tflat})
and Figure \ref{fig:opt_thick}. 
This is the highest temperature that one would infer from 
fitting the flat optically-thick disk model to the SED. 
These shielded particles are cool enough 
for sublimation not to be important --- an assumption that 
we check later.  

Particles in the disk migrate inwards due to PR drag
or aerodynamic coupling to the gaseous disk --- this migration 
is what ultimately gives rise to metal accretion onto the WD. 
As particles enter the rim and get exposed to direct starlight
their temperature rapidly goes up to 
$T_{\rm rim}=T_{\rm thin}(R_{in})\gtrsim T_{in}$ (the amount of 
energy required to heat the particle by several hundred K is
small compared to the heat of sublimation).
According to equation (\ref{eq:Tofr}) the inner rim of the 
optically thick disk lies at
\ba
R_{in}^{\rm thick}=\frac{R_\star}{2}
\left(\frac{T_\star}{T_{\rm rim}}\right)^2.
\label{eq:r_in_thick}
\ea

The fact that particles in the rim 
are illuminated by virtually unattenuated stellar radiation 
implies that the optical depth of the rim to starlight in the 
{\it radial} direction 
\ba
\tpar\approx \int\limits_0^L n(x)\times \pi a^2(x)dx
\approx 1,
\label{eq:tau=1}
\ea
where $L$ is the radial extent of the rim, $x$ is the radial 
distance away from the inner edge of the rim (i.e. the location 
where all particles sublimate; rim corresponds to $0<x<L$), 
$n(x)$ is the {\it volume number} density of particles, and we 
assume the particle cross section for starlight to be equal to 
the geometric cross section $\pi a^2$ (assuming spherical 
particles). 

Since solids lose 
mass to sublimation while drifting through the rim,
particle radius $a$ is a function of $x$; in particular
$a(x=0)=0$. The evolution of particle size due to sublimation 
is described by the following simple equation
\ba
\frac{d}{dt}\left(\frac{4\pi}{3}\rho a^3\right)\approx 
-4\pi a^2 \dot m(T_{\rm rim}),
\label{eq:mass_decrease}
\ea
which is a simplified version of equation (\ref{eq:mass_balance}) 
in which condensation has been neglected. This is a reasonable 
assumption because we will find later that the rim temperature 
$T_{\rm rim}$ is significantly higher than the quasi-static 
sublimation temperature $T_s(P_{\rm vap})$. In this case 
the flux of molecules (or atoms) 
leaving the particle surface is much higher than the flux of 
particles arriving at it (for a given surrounding vapor pressure
$P_{\rm vap}$), so that condensation can be neglected. 

We assume that the disk outside the rim 
is composed of particles of a single size $a_0$ so that $a(x=L)=a_0$.
Introducing $v_r\equiv dr/dt=dx/dt$ one can write 
$da/dt=v_r da/dx$, so that equation (\ref{eq:mass_decrease})
reduces to  
\ba
\frac{da}{dx}=\frac{\dot m(T_{\rm rim})}{\rho v_r(a,x)}.
\label{eq:dadx}
\ea
In general $v_r(a,x)$ is a function of both $a$ and $x$, see 
e.g. equation (\ref{eq:vr}) for the case of PR drag-driven 
accretion. 

We will now assume that as particles pass 
through the rim and sublimate, their 
{\it number} flux $F_N$ does not change (until they
fully sublimate) even though their
{\it mass} flux varies because their sizes go down 
as a result of sublimation. This assumption 
amounts to neglecting the possibility of particle breaking 
or merging during their travel through the rim.

Introducing the vertical thickness of the disk
$h(x)$ one can use the constancy of $F_N$ to express volume 
number density of particles in the rim $n(x)$ as
\ba
n(x)=\frac{F_N}{2\pi R_{in}}\times \frac{1}{v_r(x)h(x)}.
\label{eq:n_x}
\ea 
We can now express $v_r$ from equation (\ref{eq:dadx}),
plug it into equation (\ref{eq:n_x}) and substitute the 
resulting expression for $n(x)$ into the condition 
(\ref{eq:tau=1}) to find that 
\ba
\frac{F_N\rho}{2R_{in}\dot m(T_{\rm rim})}\int\limits_0^L
\frac{a^2}{h(x)}\frac{da}{dx}dx\approx 1.
\label{eq:cond}
\ea

To proceed further we need to make explicit 
assumptions regarding the behavior of $h(x)$.
Debris disks around WDs are expected
to be similar in properties to dense planetary rings 
around Saturn. The latter have vertical
thickness comparable to the particle size, which 
is established by collisions between particles. 
Thus, it may be natural to assume that $h(x)\sim a(x)$,
which upon plugging into equation (\ref{eq:cond})
and integrating with the condition $a(L)=a_0$ gives
\ba
\frac{F_N\rho a_0^2}{4R_{in}\dot m(T_{\rm rim})}\approx 1.
\label{eq:cond1}
\ea
The mass accretion rate of metals onto the WD 
$\dot M_Z$ is related to $F_N$ via 
$\dot M_Z=F_N \times (4\pi/3)\rho a_0^3$, so that
equation (\ref{eq:cond1}) ultimately yields
\ba
\dot m(T_{\rm rim})\approx \frac{3}{8}\frac{\dot M_Z}{2\pi R_{in} a_0}.
\label{eq:hsima}
\ea

One can try another simple approximation for the behavior of 
$h(x)$, namely assuming that $h\sim a_0=const$. In this 
case one again recovers condition (\ref{eq:hsima})
with a factor of $1/4$ instead of $3/8$. This similarity
of results suggests that, for any reasonable assumption regarding 
the behavior of $h(x)$, the condition 
\ba
\dot m(T_{\rm rim})\approx \zeta\frac{\dot M_Z}{R_{in} a_0},
\label{eq:hgen}
\ea
with $\zeta\sim 0.05-0.1$ must be satisfied in the rim. 

\begin{figure}
\plotone{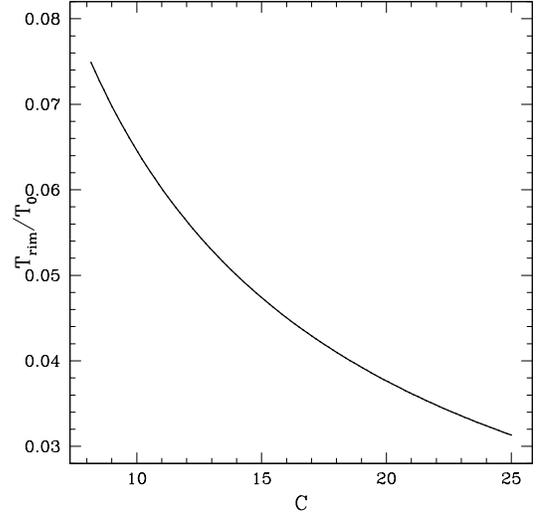}
\caption{
Solution of equation (\ref{eq:eqn2}) for the scaled temperature 
of the inner rim $T_{\rm rim}/T_0$ as a function of the 
dimensionless parameter $C$ given by equation 
(\ref{eq:C}) which contains all information about the parameters
of the system and particle properties.
\label{fig:f1}}
\end{figure}

Equation (\ref{eq:hgen}) is the condition that determines 
the value of the inner rim temperature $T_{\rm rim}$ (or, 
equivalently, the inner radius $R_{in}$) once the 
explicit form of $\dot m(T_{\rm rim})$ is specified.
Given that $2\pi R_{in} a_0$ is the area of the inner rim as
seen from the WD, equation (\ref{eq:hgen})
suggests a simple physical interpretation: the disk is
truncated at the distance $R_{in}$, where the 
full rate of sublimation from the area of its inner rim 
facing the star ($\sim 2\pi R_{in} a_0$)
roughly matches the metal accretion rate through the disk 
$\dot M_Z$. 

By taking $da\sim a_0$, $dx\sim L$ in equation (\ref{eq:dadx}) 
we estimate the time $t_{\rm cross}$ it takes particles to 
cross the rim (and sublimate): $t_{\rm cross}\sim L/v_r\sim 
\rho a_0/\dot m(T_{\rm rim})$. Using equation (\ref{eq:hgen})
to express $\dot m(T_{\rm rim})$ via $\dot M_Z$ and equation 
(\ref{eq:r_in_thick}) for $R_{in}$ we obtain
\ba
t_{\rm cross}&\sim &\frac{\rho a_0^2R_\star}{2\zeta\dot M_Z}
\left(\frac{T_\star}{T_{\rm rim}}\right)^2
\label{eq:t_cross}\\
&\approx & 400~\mbox{s}
\frac{\rho_1 a_{0,1}^2R_{\star,-2}}{\zeta_{0.1}\dot M_{Z,8}}
\left(\frac{T_\star/T_{\rm rim}}{3}\right)^2,
\nonumber
\ea
where $R_{\star,-2}\equiv R_\star/10^{-2}R_\odot$, 
$\zeta_{0.1}\equiv \zeta/0.1$. Interestingly, this estimate
is independent of the nature of the physical process driving
particle migration.   

We now plug $R_{in}$ expressed in terms of $T_{\rm rim}$ via 
equation (\ref{eq:r_in_thick}) into equation (\ref{eq:hgen}) 
to find the following transcendental equation for 
$T_{\rm rim}$ only:
\ba
\langle\alpha\rangle K_0 e^{-T_0/T_{\rm rim}}\approx 
2\zeta\frac{\dot M_Z}{R_\star a_0}
\left(\frac{T_{\rm rim}}{T_\star}\right)^2,
\label{eq:eqn0}
\ea
from which we find
\ba
\frac{T_{\rm rim}}{T_0}&=&\left(\ln\Lambda_{\rm rim}\right)^{-1},
\label{eq:eqn1}\\
\Lambda_{\rm rim}&=&\frac{\langle\alpha\rangle 
K_0 R_\star a_0}{2\zeta\dot M_Z}
\left(\frac{T_\star}{T_{\rm rim}}\right)^2,
\label{eq:Lambda_rim}
\ea
with $\Lambda_{\rm rim}\gg 1$. 
This result can be re-written in the following simple form 
amenable for iterative solution:
\ba
\frac{T_{\rm rim}}{T_0}&=&
\left[C-2\ln\frac{T_{\rm rim}}{T_0}\right]^{-1},
\label{eq:eqn2}\\
C&=&\ln\left[\frac{\langle\alpha\rangle 
K_0 R_\star a_0}{2\zeta\dot M_Z}
\left(\frac{T_\star}{T_0}\right)^2\right]
\label{eq:C}\\
&\approx &
18.6+\ln\frac{\langle\alpha\rangle_{0.1}R_{\star,-2}
a_{0,1}T_{\star,4}^2}{\zeta_{0.1}\dot M_{Z,8}}
\nonumber
\ea
where $T_{\star,4}\equiv T_\star/(10^4$ K) and the numerical 
estimate in equation (\ref{eq:C}) 
is done for olivine ($K_0=1.6\times 10^{9}$ 
g$^{-1}$ cm$^{-2}$ s$^{-1}$, $T_0=68,100$ K, see Table 
\ref{tbl:materials}). Figure \ref{fig:f1} shows the exact 
solution of equation 
(\ref{eq:eqn1}) for $T_{\rm rim}/T_0$ as a function of $C$. 
This curve is independent of the system parameters and particle 
properties ($\dot M_Z$, $a_0$, composition), which are all absorbed
into the definition of $C$. 

Using equations (\ref{eq:T_s_P}), (\ref{eq:Pvap}) and 
(\ref{eq:t_cross}) it can be trivially shown that
\ba
\frac{\Lambda_s}{\Lambda_{\rm rim}}=\frac{t_s}{t_{\rm cross}}.
\label{eq:Lambda_rat}
\ea
As a result, when the time $t_{\rm cross}$ it takes 
for a particle to cross the rim is shorter than 
the sublimation timescale $t_s$ one finds that 
$\Lambda_s\gg \Lambda_{\rm rim}$ and 
$T_{\rm rim}\gtrsim T_s$. This illustrates our expectation that
in the case of dynamical sublimation the temperature of particles 
can be higher than the quasi-static sublimation temperature
$T_s$ given by equation (\ref{eq:T_s_P}).  

For example, for the fiducial 
values of parameters adopted in equation (\ref{eq:C}) one finds 
for olivine $C=18.6$ and $T_{\rm rim}\approx 0.04T_0\approx 2700$ K,
while according to Table \ref{tbl:materials} olivine has 
$T_s\approx 2100$. The inner edge of the disk in this case is located 
very close to the WD surface, at $R_{in}\approx 7R_\star$, 
see equation (\ref{eq:r_in_thick}). 

\begin{figure}
\plotone{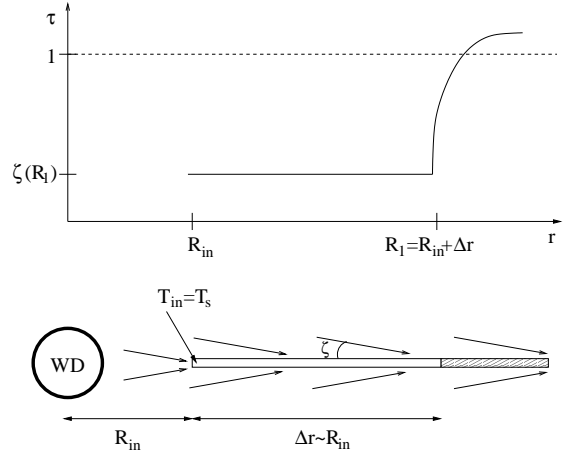}
\caption{
Schematic representation of the inner rim structure and
the surface density distribution in its vicinity in
the optically thin ($\tau_\parallel\lesssim 1$) inner 
disk. The inner optically thin part of the disk (unshaded) 
is directly illuminated by the star over a broad range 
of radii $\Delta r\sim R_{in}$. The optically thick part
(shaded) starts at $R_{1}=R_{in}+\Delta r$. Compare with 
Figure \ref{fig:opt_thick}.
\label{fig:opt_thin}}
\end{figure}

At the same time, just behind the rim the disk temperature is 
$T_{in}=T_{\rm thick}(R_{in})\approx 1600$ K $<T_s$ for 
$T_\star=10^4$ K, see equation (\ref{eq:Trat}). This 
verifies our previous assumption of relatively low 
particle temperature (i.e. $T_{in}<T_s$) just behind 
the rim, justifying the disregard of particle sublimation 
in this region. On the other hand, this value of $T_{in}$
is clearly high enough for the disk to produce enough 
near-IR emission corresponding to $T\sim 1500$ K, in 
agreement with observations.

Equation (\ref{eq:Lambda_rat}) also emphasizes the necessity of 
particle accretion for maintaining the superheated
inner rim: if $\dot M_Z\to 0$ then according to equation 
(\ref{eq:t_cross}) $t_{\rm cross}\to \infty$ and the hot 
inner rim does not exist.


\section{Sublimation radii in disks with optically 
thin ($\tpar\lesssim 1$) inner regions}
\label{sect:opt_thin}

We now look at the case of a disk, the inner part of
which is {\it optically thin} for incident stellar radiation, i.e.
$\tpar\lesssim 1$, see Figure \ref{fig:opt_thin}. 
As demonstrated by \citet{bochkarev_2011} such 
situation naturally arises for 
a low mass disk, which starts with $\tpar\lesssim 1$ 
everywhere, or for a moderately massive disk, which hasn't
gone through the runaway accretion phase. In the latter 
case, as shown in \citet{bochkarev_2011}, an optically 
thin tail of solid material 
with $\tau_\parallel\sim 1$ (or $\tau\sim \zeta\ll 1$) 
naturally develops as an inward extension of the optically thick 
part of the disk under the action of the PR drag. 

Particles in such optically thin tail are directly exposed
to starlight, meaning that their equilibrium temperature 
is given by equation (\ref{eq:Tofr}). Also, \citet{rafikov_sep2011} 
and \citet{metzger_2012} have shown that the dynamics of these 
optically thin regions, including the radial drift of 
particles, is determined primarily by PR drag. Then
the radial migration speed is just
\ba
v_{r,PR}=\frac{3}{8\pi}\frac{L_\star}{\rho a_0 c^2}\frac{1}{r}.
\label{eq:vr}
\ea
The characteristic timescale $t_{\rm PR}\equiv r/v_{r,PR}$ on which 
the particle distance and temperature vary under the
action of the PR drag is then
\ba
t_{\rm PR}=\frac{8\pi}{3}\frac{\rho a_0 c^2}{L_\star}r^2\approx
10^4\mbox{yr}\frac{\rho_1 a_{0,1}}{L_{\star,-3}}
\left(\frac{r}{0.2R_\odot}\right)^2.
\label{eq:t_PR}
\ea

Since $t_{\rm PR}$ is much longer than the sublimation 
timescale $t_s$ given by equation (\ref{eq:subl_time}) it
is clear that in the optically thin disks sublimation must
be occurring in a quasi-static fashion: particles slowly 
drift inward under the action of the PR drag and their 
temperature steadily rises. At some radius $R_{in}^{\rm thin}$  
their temperature reaches $T_s$, and particles turn into metal gas on a 
(short) sublimation timescale $t_s$. That means that the
inner edge of the optically thin disk is set by the 
condition $T_{\rm thin}(R_{in}^{\rm thin})=T_s$, with 
$T_s$ given by equation (\ref{eq:T_s_P}). Thus,
\ba
R_{in}^{\rm thin}=\frac{R_\star}{2}
\left(\frac{T_\star}{T_s}\right)^2.
\label{eq:r_in_thin}
\ea
In particular, according to Table 
\ref{tbl:materials} we need to take $T_s\approx 2100$ K 
for olivine, which when plugged in the equation 
(\ref{eq:r_in_thin})
yields $R_{in}\approx 11R_\star$ for $T_\star=10^4$ K. 
This is about $60\%$ further from the star that in the 
case of an optically thick disk, see \S \ref{sect:thick}.


\section{Application to observed systems.}
\label{sect:apps}

We now apply ideas developed in \S \ref{sect:thick} 
to a sample of observed WDs with debris disks. We 
start by rewriting the expression (\ref{eq:C}) for $C$
as
\ba
C&=&C_\star+C_p,
\label{eq:C_split}\\
C_\star &\equiv & 
\ln\left[\frac{R_\star T_\star^2}{\dot M_Z}
\right],~~~
C_p \equiv\ln\left[\frac{\langle\alpha\rangle 
K_0 a_0}{2\zeta T_0^2}\right].
\label{eq:Cs}
\ea
Here $C_\star$ is a parameter, which depends only on 
measurable properties of the system --- WD radius, effective 
temperature, and metal accretion rate. All parameters
characterizing the particle properties --- $K_0$, $a_0$, etc. ---
are absorbed into $C_p$. Assuming a particular composition of
particles and a value of particle radius $a_0$ fixes $C_p$
and allows one to obtain a theoretical relation between
$T_{\rm rim}$ and $C_\star$ using equations (\ref{eq:eqn1}),
(\ref{eq:C_split}), and (\ref{eq:Cs}). By looking at different
particle compositions one can compare the corresponding 
theoretical $T_{\rm rim}(C_\star)$ curves with the 
properties of observed systems.

\begin{center}
  \begin{deluxetable*}{lccccccccccc}
    \tablecolumns{11}
    \tablecaption{Properties of disk-hosting WDs used in this work
    \label{tbl:systems}}
    \tablehead{
      \colhead{Name} &
      \colhead{SpT} &
      \colhead{$M_{\star}$} &
      \colhead{$R_{\star}$} &
      \colhead{$T_{\star}$} &
      \colhead{$\log_{10} \dot{M}$} &
      \colhead{Gas Disk} &
      \colhead{$T_{in}$} &
      \colhead{$T_{\rm rim }$} &
      \colhead{$C_{\star}$} &
      \colhead{Ref.} \\
      \colhead{} &
      \colhead{} &
      \colhead{$M_{\sun}$} &
      \colhead{$R_{\sun}$} &
      \colhead{K} &
      \colhead{g/s} &
      \colhead{Detected} &
      \colhead{K} &
      \colhead{K} &
      \colhead{CGS} &
      \colhead{}
      }
    \startdata
      GD 16 & DAZB & 0.59 & 0.014 & 11500 & 8.0 &  &
      \hspace{6pt}1300\tna & 2460 & 21.0 & 1\\
      GD 133 & DAZ & 0.59 & 0.014 & 12200 & 8.5 &  & 1200 & 2380 & 19.9 &
      1,2\\
      GD 40 & DBZ & 0.59 & 0.013 & 15200 & 9.9 &  & 1200 & 2560 & 17.1 & 1\\
      GD 56 & DAZ & 0.60 & 0.015 & 14200 & 8.5 &  & 1700 & 3160 & 20.3 &
      1,2\\
      J1228+1040 & DAZ & 0.77 & 0.011 & 22020 & 9.3 & yes & 1670 & 3610 & 19.0
      & 3,4\\
      PG1015+161 & DAZ & 0.61 & 0.014 & 19300 & 9.3 &  & 1200 & 2770 & 19.0 &
      1,2\\
      Ton345 & DBZ & 0.70 & 0.010 & 18600 & 9.4 & yes & 1500 & 3180 & 18.4 &
      5,6,7\\
      SDSS1043+0855 & DAZ & 0.66 & 0.012 & 17900 & 9.0 & yes & 1400 & 3000
      & 19.4 & 8\\
      G29-38 & DAZ & 0.62 & 0.013 & 11700 & 8.7 &  & 1200 & 2350 & 19.3 &
      1,9,10\\
      GD 362 & DAZB & 0.73 & 0.013 & 10500 & 10.4 &  & 1200 & 2260 & 15.2 &
      1,9,11\\
      SDSS0959 & DAZ & 0.64 & 0.012 & 13280 & 7.9 &  & 1600 & 2970 & 21.3 &
      12\\
      SDSS1221 & DAZ & 0.73 & 0.011 & 12250 & 7.7 &  & 1400 & 2640 & 21.6 &
      12\\
      SDSS1557 & DAZ & 0.42 & 0.018 & 22810 & 8.8 &  & 1400 & 3250 & 20.8 &
      12\\
      GD 61 & DBZ & 0.71 & 0.011 & 17280 & 8.81 &  & 1300 & 2820 & 19.7 &
      13,14\\
      J0738+1835 & DBZ & 0.84 & 0.010 & 13950 & 11.11  & yes & 1600 & 3020
      & 13.9 & 16\\
      HE 0110-5630 & DBAZ & 0.71 & 0.012 & 19200 & 8.4 &  & 1000 & 2450 & 20.9
      & 17,18\\
      HE 1349-2305 & DBAZ & 0.67 & 0.012 & 18200 & 8.7 &  & 1700 & 3430 & 20.1
      & 17,18\\
    \enddata
    \tablenotetext{a}{These numbers were calculated in this paper.}
    \tablenotetext{b}{(1) \citealt{farihi_2009}; (2) \citealt{jura_jul2007}; 
    (3) \citealt{brinkworth_2009}; (4) \citealt{gansicke_2006}; (5)
    \citealt{farihi_2010}; (6) \citealt{melis_2010}; (7)
    \citealt{gansicke_2008}; (8) \citealt{brinkworth_2012}; (9)
    \citealt{farihi_2008}; (10) \citealt{zuckerman_2003}; (11)
    \citealt{zuckerman_2007}; (12) \citealt{farihi_2012}; (13)
    \citealt{farihi_feb2011}; (14) \citealt{jura_2012}; (15)
    \citealt{kilic_2012}; (16) \citealt{dufour_2012}; (17)
    \citealt{girven_2012}; (18) \citealt{koester_2005}.}
  \end{deluxetable*}
\end{center}
 
Such comparison requires the knowledge of $R_\star$, 
$T_\star$, $\dot M_Z$, which we take from the 
literature. One also needs to know $T_{\rm rim}$ for
each of the WD+disk systems, and we derive this parameter
as $T_{\rm thin}$ from equation (\ref{eq:Trat}), in 
which we use $T_{in}$ --- the innermost disk 
temperature inferred from the SED fitting --- for 
$T_{\rm thick}$. We use the 
values of $T_{in}$ determined in the literature when 
available, and we provide our own fits otherwise. 
The summary of WD+disk 
parameters used in our comparison with theory is 
provided in Table \ref{tbl:systems}.

\begin{figure}
\plotone{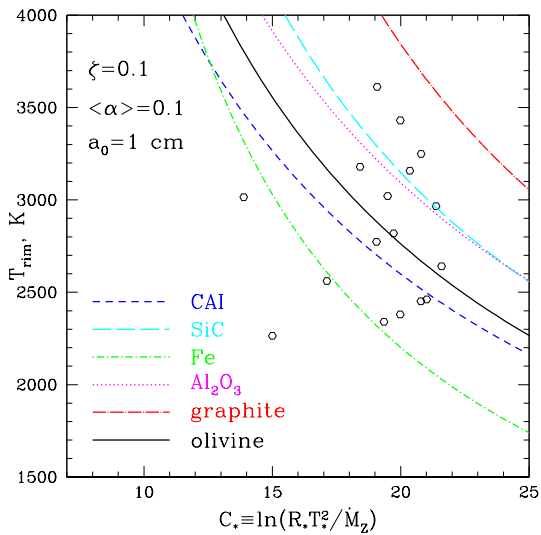}
\caption{
Comparison of observed WD+disk properties with theoretical
predictions in $C_\star-T_{\rm rim}$ space ($C_\star$ is
defined by equation (\ref{eq:Cs})). Theoretical 
$T_{\rm rim}(C_\star)$ curves computed for different particle 
compositions are labeled on the plot. Their calculation 
assumes $a_0=1$ cm particles, $\zeta=0.1$, and $\accom=0.1$. 
Note that most of the observed systems are consistent 
with particles being made of Si-bearing materials, such as
olivine or CAI.
\label{fig:comparison}}
\end{figure}

In Figure \ref{fig:comparison} we show theoretical 
$T_{\rm rim}(C_\star)$ curves for different particle 
compositions. In our calculations we always
assume $a_0=1$ cm particles, $\accom=0.1$, and $\zeta=0.1$
(all dimensional quantities are expressed in CGS units). 
We also plot the locations of observed 
systems from Table \ref{tbl:systems} in $C_\star-T_{\rm rim}$ 
space with hexagons.

As expected, very refractory particles made of graphite,
SiC, and Al$_2$O$_3$ are characterized by considerably higher 
values of theoretical $T_{\rm rim}$ (for the same $C_\star$) 
than if they were to have more volatile compositions, e.g. 
were made of iron. The difference in $T_{\rm rim}$ can 
easily exceed $10^3$ K.

The vast majority of observed systems lies in between the two
extremes determined by the iron and graphite. It is clear
from this plot that the pure graphite composition is not 
acceptable for particles in the observed WD+disk systems ---
all of them are below the corresponding theoretical curve.   
Also, only a handful of systems lie close to the theoretical
$T_{\rm rim}(C_\star)$ curve for iron. The majority 
of observed WD+disk systems tend to gravitate towards 
$T_{\rm rim}(C_\star)$ curves computed for CAI and olivine-like
compositions. At the same time about a third of the systems 
in the upper right corner of the figure are consistent with 
more refractory compositions such as SiC or Al$_2$O$_3$.

When comparing characteristics of observed systems with 
theoretical predictions for $T_{\rm rim}(C_\star)$, a couple of 
issues have to be kept in mind. First, observational 
determination of parameters of the WD+disk systems is
prone to errors. This is not so serious for the 
determination of $T_\star$, which is typically quite 
accurate, or $R_\star$, which does not span a large range 
anyway. However, the determination of $\dot M_Z$ from the data 
depends on the unknown composition of the parent body 
that formed the disk, and may have large error bars.
On the other hand, $C_\star$ depends on these characteristics
only logarithmically, so that even large uncertainties in these
parameters would result in a relatively small horizontal shift
of observational points in Figure \ref{fig:comparison}. 

The uncertainty in measuring $T_{\rm rim}$ is much more
serious. This is because the determination of $T_{in}$
relies on fitting the flat disk model to the SED, and  $T_{in}$ can be
highly degenerate 
with other parameters, such as the disk inclination
\citep{girven_2012}. Also, 
according to equation (\ref{eq:Trat}) 
$T_{\rm rim}\propto T_{in}^{2/3}$, so that the errors
in determination of $T_{in}$ from SED directly propagate
into the uncertainty in $T_{\rm rim}$. As a result,
observational data points in Figure \ref{fig:comparison}
can have significant vertical errorbars.

Another thing to keep in mind, is that when computing 
the theoretical $T_{\rm rim}(C_\star)$
curves we make certain assumptions about particle properties,
such as their size $a_0$ or accommodation coefficient 
$\accom$. Variation of these parameters from their adopted
values affects the value of $C_p$ and causes horizontal shift 
of the $T_{\rm rim}(C_\star)$ curves. For example, increasing 
the value of accommodation coefficient $\accom$ from $0.1$
to $1$ displaces the theoretical curves to the left by 
$\Delta C_\star=2.3$. This would put observational 
data points in better agreement with the more refractory 
particle compositions.


\section{Discussion.}
\label{sect:disc}

The physical model for the inner rim structure in the
optically thick case presented in 
\S \ref{sect:thick} naturally allows us to explain the 
high inner disk temperatures $T_{in}$ inferred from the 
SEDs of debris disks around some WDs. The existence of
a narrow inner rim of the disk heated to a temperature 
$T_{\rm rim}$ 
above the quasi-static sublimation temperature $T_s$
(see equation (\ref{eq:T_s_P})) is the key ingredient 
of the model.

The radial width of the inner rim $L$ can be estimated by
multiplying the time to cross it $t_{cross}$ by the velocity 
$v_{r,PR}$, given by equations 
(\ref{eq:t_cross}) and (\ref{eq:vr}) correspondingly:
\ba
L\sim 10~\mbox{cm} \frac{a_{0,1}R_{\star,-2}L_{\star,-3}}
{\zeta_{0.1}\dot M_{Z,8}}\left(\frac{0.2\mbox{AU}}{R_{in}}\right)
\label{eq:L}
\ea
Thus, one typically finds the width of the inner rim to be
$\sim 10$ particle radii.

Note that the radial speed of particles in massive disks 
can be affected by aerodynamic coupling between the particulate 
and gaseous disks, and in consequence deviate 
from $v_{r,PR}$. Nevertheless,
equation (\ref{eq:L}) serves as a reasonable order of magnitude 
estimate of $L$ and clearly demonstrates that $L\ll R_{in}$. 
As a result, the contribution of the hot inner rim to the SED
of the debris disk is completely negligible, and its 
spectrum is determined only by emission from the parts of the 
disk located behind the rim.  

\begin{figure}
\centering
\includegraphics[width=1.0\linewidth, keepaspectratio=true]{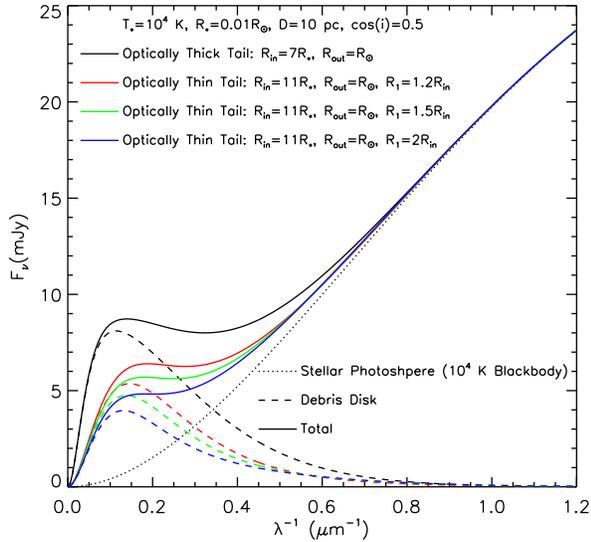}
\caption{
Spectra of debris disks with optically thick and optically thin
inner regions. All models feature optically thick regions 
with $\tau=10$, which extend from $R_{out}=R_\odot$ to 
$R_{in}^{\rm thick}$ in the optically thick case and to $r=R_1$
(indicated in the panel) in the models with optically thin tails.
Inside $R_1$ tails have $\tau=\zeta(R_1)\ll 1$. Stellar parameters
are indicated on the panel. See text for more details. 
\label{fig:specs}}
\end{figure}

Our results in \S \ref{sect:thick}-\ref{sect:opt_thin} allow
us to address the differences in spectra of disks with
optically thick or thin inner regions. In Figure \ref{fig:specs}
we show several spectra produced by disks around a $T_\star=10^4$ K, 
$R_\star=0.01R_\odot$ WD located 10 pc away from us, and inclined 
with respect to our line of sight with $\cos i=0.5$. The model,
which is optically thick everywhere, has constant optical 
depth $\tau=10$ and extends from the outer radius 
$R_{out}=R_\odot$ to $R_{in}^{\rm thick}\approx 7R_\star$ 
given by equation (\ref{eq:r_in_thick}). Models with optically
thin tails also have constant optical depth $\tau=10$ between
$R_{out}=R_\odot$ and some intermediate radius $R_1$, which 
is different for each model. Inside of $R_1$ we assume
an optically thin tail with constant $\tau=\zeta(R_1)$ 
(or $\tau_\parallel=r/R_1$) to extend from $R_1$ down to 
$R_{in}^{\rm thin}\approx 11R_\star$ given by equation 
(\ref{eq:r_in_thin}). This is the characteristic distribution 
of $\tau$ in the inner optically thin tails of the disks
evolving under the action of the PR drag, see 
\citet{bochkarev_2011}. 

One can see that the disk which is optically thick 
everywhere produces more flux. This is expected because 
disks with optically thin tails do not extend as far 
inward, and are inefficient at absorbing and re-radiating in regions
interior to $r=R_1$. The spectral shape is also different,
in part because particles in the optically thin tail
are hotter than particles in the optically thick 
tail at the same radius. This
may allow one to diagnose the presence of an inner
optically thin tail using just the disk SED. Such optically
thin tails may be expected in systems characterized by 
$\dot M_Z\sim 10^8$ g s$^{-1}$, i.e. close to the value 
provided by PR drag alone. In the runaway scenario
of \citet{metzger_2012}, one expects systems with higher
$\dot M_Z$ to be evolving due to aerodynamic coupling 
with the gaseous disk, in which case the disk is 
optically thick all the way down to $R_{in}^{\rm thick}$. 

Comparison of our theory with characteristics of observed
WD+debris disk systems shows that in general (barring the 
uncertainties related to measurement errors and 
poorly constrained modeling parameters) properties of 
these systems are consistent with Si-rich particle 
composition. In other words, we find that CAI- or
olivine-like compositions of particles are in reasonable 
agreement with the locations of the inner rims in the 
majority of observed disk-hosting systems. 

This result reinforces previous conclusions about the 
Si-rich nature of the accreted material based on different 
and independent lines of evidence. In particular (and most 
importantly), direct measurements of the metal abundances 
in the WD atmospheres show that the composition of accreted 
material is consistent with that of the inner Solar System 
bodies, which are known to be Si-rich (\citealt{zuckerman_2007};
\citealt{klein_2010}, 2011; \citealt{jura_may2012}). These measurements 
also demonstrate the accreted bodies to be carbon-poor
\citep{jura_2006}, which is again consistent with our results 
--- essentially none of the observed WD+disk systems lie 
close to the C-based curve in Figure \ref{fig:comparison}.
Additional evidence in favor of Si-rich particle
composition comes from the measurement of 10-$\mu$m bump
in debris disk spectra obtained with {\it Spitzer IRS}
(\citealt{jura_may2007}, 2009). This feature is usually interpreted 
as being produced by the $\mu$m-size silicate particles.

\begin{figure}
\plotone{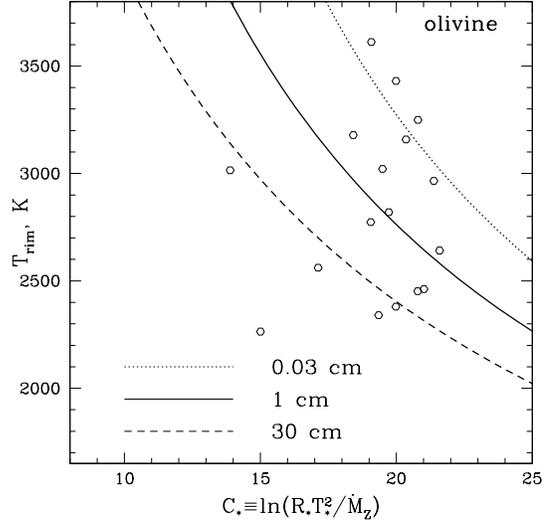}
\caption{
Effect of varying the particle radius $a$ on the theoretical
$T_{\rm rim}-C_\star$ dependence for olivine and 
comparison with observational data. 
\label{fig:var_rad}}
\end{figure}

Using these independent lines of evidence supporting the 
Si-rich nature of the debris disk constituents we may 
approach our findings from a different perspective. In 
particular, by postulating disk particles to be Si-rich 
we can put constraint on their sizes. Results presented
in Figure \ref{fig:comparison} do a reasonably good 
job at reproducing characteristics of observed systems  
by assuming $a=1$ cm particles. Varying $a$ would 
displace the theoretical curves horizontally and they would 
remain consistent with observations only within a certain 
range of particle sizes. Figure \ref{fig:var_rad}
illustrates this variation of the $T_{\rm rim}-C_\star$
relation; one can easily infer from it that particle 
sizes should lie within the range $a=0.03-30$ cm. 
Otherwise the properties of the inner disk rims in the majority 
of the observed systems will not be consistent 
with our theoretical calculations. 

Interestingly, this range of particle sizes is consistent
with other indirect measurements of $a$ reported in the 
literature. In particular, \citet{graham_1990} found 
$a\lesssim 10$ cm based on the variability of the 
reprocessed IR emission of disk particles. \citet{metzger_2012} found
$a\lesssim$ several cm to provide the best fit to the
runaway picture of the disk evolution. Finally, 
Saturn rings, which are thought to be rather close in 
properties to circum-WD debris disks, are 
also predominantly composed of $1-100$ cm particles
\citep{cuzzi_2010}.

Our model naturally explains the presence of solid particles  
even around hot WDs, with $T_\star\approx 20,000$ K, 
e.g. J1228+1040 and SDSS1557. Conventional theory 
finds it difficult to account for such systems.
Indeed, equation (\ref{eq:Tofr}) predicts that around 
$T_\star=20,000$ K,
$R_\star=0.015 R_\odot$ WD directly illuminated 
particles must have a temperature of $1700$ K at the 
tidal radius of $\sim R_\odot$. This is significantly higher
than the sublimation temperature of $1300-1500$ K
usually assumed based on protoplanetary disk studies 
\citep{lodders_2003}. Our calculations first show that in fact 
the sublimation temperature $T_s$ can easily be higher 
than 1700 K, see Table \ref{tbl:materials}, which 
guarantees the survival
of even the optically thin disks with directly exposed 
particles within tidal radii of hot WDs. Second, in 
the optically thick case, shielding of the disk by 
the inner rim particles allows $R_{in}$ 
to be as small as $0.3R_\odot$ (for $R_\star=10^{-2}R_\odot$,
and keeping all other parameters equal to their values in 
equation (\ref{eq:C})).

Our present calculations were designed to demonstrate the 
main qualitative features of the inner rim structure and
thus made a number of simplifying assumptions. One of them 
is the single size of particles in the disk, while in reality 
a distribution of particle sizes should be present. We expect 
that in this case the value of $a$ in the definition 
(\ref{eq:Cs}) of $C_p$ would be replaced with some properly 
weighted average of the particle size distribution, but 
the main results would not change.

Another simplification is the assumed single chemical 
composition of all particles. If the disk contains particles 
of different compositions, with different $K_0$ and $T_0$,
then one may expect a ``multi-rim'' structure to form, in
which different chemical species sublimate at different 
radii. In this case the inner radius of the disk would be 
determined by the properties of the {\it most refractory} 
particles in the disk that survive at the closest separation 
from the WD. Observations of the inner disk properties 
(i.e. $T_{in}$) would then be sensitive to characteristics 
of only this particular particle population (as long as 
the disk is optically thick everywhere).


\section{Summary.}
\label{sect:sum}

We explored the structure of the inner parts of compact 
debris disks around WDs with the goal of resolving the
``inner rim puzzle'' --- the difficulty with reconciling 
the high inner disk temperatures inferred from the SED with 
the material properties of putative constituent particles. 
We first show that because of the much higher vapor pressure
of metals in these hydrogen-poor disks compared 
to the hydrogen-rich protoplanetary disks, the quasi-static 
sublimation temperature $T_s$ of different species in 
circum-WD disks is typically 300-400 K higher
than in their conventional protoplanetary analogues. This
revised value of $T_s$ determines the (smaller than was 
thought before) value of the inner radius for the optically 
thin disks, given by equations (\ref{eq:T_s_P}) \& 
(\ref{eq:r_in_thin}).

We demonstrate that optically thick 
circum-WD disks feature narrow inner rims, which are 
superheated above $T_s$. This allows inner disk radii
in such systems, described by equations (\ref{eq:r_in_thick}), 
(\ref{eq:eqn1}), \& (\ref{eq:Lambda_rim}), to lie quite 
close to the WD, easily at 
separations $\sim 10R_\star$. The main physical ingredients 
needed for the existence of such superheated inner rim are
(1) accretion of particles through the disk, which can be 
easily maintained at the necessary level by 
Poynting-Robertson drag, (2) shielding of particles behind the
rim from starlight by the rim particles, and (3) dynamic nature
of the sublimation process inside the hot rim.
The combination of these ingredients naturally allows particles 
to reach temperatures of order $1600-1700$ K just behind 
the rim, which is needed to explain the SEDs of some systems. 
Particles inside the rim are heated to $2500-3500$ K
and undergo rapid sublimation as they migrate in. Using
this model we can naturally explain the existence of 
particulate debris disks even around hot WDs, with effective
temperature $\gtrsim 20,000$ K.

We compare our predictions with existing observations of the 
WD+disk systems. We find that properties of particles in
debris disks are consistent with Si-rich composition,
such as olivine or CAI-like material. Very refractory 
(such as graphite) or more volatile (such as iron) compositions
are clearly disfavored by this comparison. Assuming that 
circum-WD disks are indeed composed of Si-rich particles
we constrain typical particle size to lie roughly between 
$0.03$ and $30$ cm, in agreement with other indirect evidence 
for cm-size objects in such disks. 

\acknowledgements

The authors thank Bruce Draine and Michael Jura for stimulating
discussions. The financial support for this work is provided 
by the Sloan Foundation and NASA via grant NNX08AH87G. 

\nocite{klein_2011}
\nocite{kilic_2006}
\nocite{gansicke_2007}



\appendix

\section{Thermal balance in the disk of solids.}
\label{sect:thermal}

Grains in the inner rim are being heated and 
cooled by four processes: (1) heating by starlight, 
(2) heating by gas, (2) cooling by thermal radiation
from particle surfaces, and (4) removal of thermal energy 
by sublimating atoms/molecules. All these processes 
scale linearly with the surface area of the particles. 
As a result, the temperature of grains directly 
exposed to starlight (assuming full absorption of the incoming 
radiation) is implicitly given as a function of the distance 
from the WD by the following formula \citep{podolak_2010}:
\ba
&& \frac{1}{4}\sigma T_\star^4\left(\frac{R_\star}{r}\right)^2+Q_{\rm gas} =
\sigma T^4+Q_{\rm sub},
\label{eq:heat_eq}\\
&& Q_{\rm sub}=\dot m(T)L_{\rm sub},~~~~~
Q_{\rm gas}=\frac{\rho_g c_{\rm s}}{2\mu} \varepsilon k_{B}
\left(T_{\rm gas}-T \right),
\label{eq:gas_heating}
\ea
where $L_{\rm sub}$ is the specific heat of sublimation of
the particle material, $\varepsilon$ is the efficiency of
heat exchange between gas and particles, and $\rho_g$ is 
the gas density. 

Using equation (\ref{eq:Mdot}) we can estimate
$\Sigma_g\approx \dot M_Z/(3\pi\nu)$ so that
\ba
\rho_g=\frac{\Sigma_g\Omega}{c_s}=\frac{\dot M_Z\Omega^2}
{3\pi\alpha_\nu c_s^3}.
\label{eq:rho_g}
\ea
This allows us to compare the contribution of gas heating 
$Q_{\rm gas}$ with stellar irradiation $Q_\star$ 
(first term in the left hand side of equation (\ref{eq:heat_eq})):
\ba
\frac{Q_{\rm gas}}{Q_{\star}} &\approx&
\frac{2\varepsilon}{3\pi}\frac{GM_{\star}\dot{M}_{Z}}
{\alpha_{\nu}r R_{\star}^{2}{\sigma}T_{\star}^{4}}
\left(1-\frac{T}{T_{\rm gas}}\right)
\sim 10^{-4}\frac{\varepsilon\dot M_{Z,8}M_{\star,1}}
{\alpha_{\nu,-2}T_{\star,4}^4 R_{\star,-2}^2}
\left(\frac{r}{0.2R_{\odot}}\right)^{-1}.
\ea
Therefore, gas heating is typically unimportant 
for the thermal balance of particles compared to 
heating by starlight, in contrast to the conclusion
reached by \citet{jura_may2007}, who looked at 
conduction in gas phase as the means to lower 
particle temperature. This difference is predominantly 
caused by the high gas density ($\sim 10^2$ times higher 
than in equation (\ref{eq:rho_g})) used in \citet{jura_may2007}.

Using prescription (\ref{eq:mdot}) with $\accom=0.1$ and 
the typical (for olivine) value of 
$L_{\rm sub}=3.2\times 10^{10}$ erg g$^{-1}$ from 
\citet{kimura_2002} we can also estimate the relative
contribution of sublimation to the cooling of 
particles by evaluating $Q_{\rm sub}/\sigma T^4$.
We find this ratio to be about unity for particles heated 
to $\approx 3400$ K. At $T=3000$ K the ratio of the energy 
loss by sublimation to radiative cooling rate is about 
$0.1$. Thus, for WD+disk systems with $T_{\rm rim}\lesssim 
3000$ K one can safely neglect $Q_{\rm sub}$ in equation 
(\ref{eq:heat_eq}). Then the thermal balance everywhere in 
the disk is determined by the equilibrium between 
stellar heating and radiative cooling only, which provides
justification for using equations (\ref{eq:Tofr}) and 
(\ref{eq:Tflat}) in this work. This assumption is good 
for the majority of observed systems shown in Figure 
\ref{fig:comparison}, and even for a handful of systems
with $T_{\rm rim}\approx 3000-3500$ our theoretical curves 
should still be at least qualitatively correct.


\section{Data on the mass sublimation rates}
\label{app:sublimation}

Here we present the details on the derivation of mass sublimation
rates for different elements shown in Table \ref{tbl:materials}.

{\bf Olivines} 
Calculation of the vapor pressure for the olivine-like 
silicate grains (e.g. Mg$_2$SiO$_4$) is complicated due to the fact that
these molecules do not exist in the gas phase. Nevertheless, 
\citet{guhathakurta_1989} suggest the following expression for 
the (number) rate of Si sublimation from the olivine surface:
$R_{\rm Si}\approx 7\times 10^{30}\accom \exp(-68,100/T)$ 
cm$^{-2}$ s$^{-1}$. The mass sublimation rate of olivine is then  
given by $\mu_{\rm oli}R_{\rm Si}$, where $\mu_{oli}=141~m_p$ is
the mean molecular weight of Mg$_2$SiO$_4$.

{\bf Graphite} 
For pure graphite \citet{guhathakurta_1989}
give the (number) rate of C sublimation from the graphite 
surface of $R_{\rm C}\approx 4.6\times 10^{30}\accom \exp(-81,200/T)$,
which then allows us to calculate $K_0$ from the mass sublimation
rate $R_{\rm C}\mu_{\rm C}$, where $\mu_{\rm C}=12~m_p$ is
the mean molecular weight of carbon.

{\bf CAI} 
\citet{richter_2007} consider evaporation of CAI-like
liquids and come up with the following (number) rate of Si
escaping a CAI-like surface: $R_{\rm Si}\approx 2.5\times 10^{31}
\accom \exp(-69,400/T)$. Using gehlenite 
(Ca$_2$Al$_2$SiO$_7$) as a typical CAI-like material (mean
molecular weight 274 $m_p$) we obtain sublimation parameters
indicated in Table \ref{tbl:materials}.

{\bf Iron} 
\citet{zatisev_2001} provide the data on the vapor
pressure of Fe: $P^{\rm Fe}_{\rm vap}=2.8\times 10^{11}\exp(-45,400/T)$ 
Pa for $T\approx 1800-1900$ K. From these data we determine the mass
sublimation rate according to the formula $\dot m_{\rm Fe}=\accom
P^{\rm Fe}_{\rm vap}\left(\mu_{\rm Fe}/2\pi k_B T\right)^{1/2}$, where
$\mu_{\rm Fe}\approx 56 m_p$ and 
we take $T=1600$ K with the expectation that the thermophysical 
parameters of Fe remain roughly the same at this temperature.  

{\bf SiC} Using the thermophysical data presented in \citet{chase_1985} we
derive the following fit to the behavior of the vapor
pressure of Si above the SiC surface in the range $T=1800-3000$ K: 
$P^{\rm Si}_{\rm vap}=9\times 10^{13}\exp(-73,700/T)$ dyne cm$^{-2}$.
Since the surface loses C atoms in addition to Si we evaluate 
SiC mass loss rate as $\dot m_{\rm SiC}=
\accom P^{\rm Si}_{\rm vap}\left(\mu_{\rm SiC}/\mu_{\rm Si}\right)
\left(\mu_{\rm Si}/2\pi k_B T\right)^{1/2}$, where $\mu_{\rm SiC}=
40m_p$, $\mu_{\rm Si}=28m_p$ and we take $T=2400$ K.

{\bf Al$_2$O$_3$} For pure corundum (Al$_2$O$_3$) the basic reaction 
which is thermodynamically most likely is Al$_2$O$_3\to 2$AlO$+$O 
(a different reaction dominates for Al-Al$_2$O$_3$ mixture, see
\citealt{brewer_1951}). Using the data in \citet{chase_1985} we find that 
the vapor pressure of O above 
the corundum surface is  $P^{O}_{\rm vap}=
3.6\times 10^{14}\exp(-80,500/T)$ dyne cm$^{-2}$ for $T=1800-3000$ K.
Accounting for the mass of Al leaving the surface together with O
we arrive at the sublimation characteristics of corundum presented
in Table \ref{tbl:materials}.


\begin{thebibliography}{49}
\expandafter\ifx\csname natexlab\endcsname\relax\def\natexlab#1{#1}\fi

\bibitem[{{Alcock} {et~al.}(1986){Alcock}, {Fristrom}, \&
  {Siegelman}}]{alcock_1986}
{Alcock}, C., {Fristrom}, C.~C., \& {Siegelman}, R. 1986, ApJ, 302, 462

\bibitem[{{Bochkarev} \& {Rafikov}(2011)}]{bochkarev_2011}
{Bochkarev}, K.~V., \& {Rafikov}, R.~R. 2011, ApJ, 741, 36

\bibitem[{Brewer \& Searcy(1951)}]{brewer_1951}
Brewer, L., \& Searcy, A.~W. 1951, Journal of the American Chemical Society,
  73, 5308

\bibitem[{{Brinkworth} {et~al.}(2012){Brinkworth}, {G{\"a}nsicke}, {Girven},
  {Hoard}, {Marsh}, {Parsons}, \& {Koester}}]{brinkworth_2012}
{Brinkworth}, C.~S., {G{\"a}nsicke}, B.~T., {Girven}, J.~M., {et~al.} 2012,
  ApJ, 750, 86

\bibitem[{{Brinkworth} {et~al.}(2009){Brinkworth}, {G{\"a}nsicke}, {Marsh},
  {Hoard}, \& {Tappert}}]{brinkworth_2009}
{Brinkworth}, C.~S., {G{\"a}nsicke}, B.~T., {Marsh}, T.~R., {Hoard}, D.~W., \&
  {Tappert}, C. 2009, ApJ, 696, 1402

\bibitem[{{Chase} {et~al.}(1985){Chase}, {Davies}, {Downey}, {Frurip},
  {McDonald}, \& {Syverud}}]{chase_1985}
{Chase}, M.~W.~J., {Davies}, C.~A., {Downey}, J.~R.~J., {et~al.} 1985, J. Phys.
  Chem. Ref. Data, 14, Suppl. No. 1

\bibitem[{{Chiang} \& {Goldreich}(1997)}]{chiang_1997}
{Chiang}, E.~I., \& {Goldreich}, P. 1997, ApJ, 490, 368

\bibitem[{{Cuzzi} {et~al.}(2010){Cuzzi}, {Burns}, {Charnoz}, {Clark},
  {Colwell}, {Dones}, {Esposito}, {Filacchione}, {French}, {Hedman}, {Kempf},
  {Marouf}, {Murray}, {Nicholson}, {Porco}, {Schmidt}, {Showalter}, {Spilker},
  {Spitale}, {Srama}, {Srem{\v c}evi{\'c}}, {Tiscareno}, \&
  {Weiss}}]{cuzzi_2010}
{Cuzzi}, J.~N., {Burns}, J.~A., {Charnoz}, S., {et~al.} 2010, Science, 327,
  1470

\bibitem[{{Debes} \& {Sigurdsson}(2002)}]{debes_2002}
{Debes}, J.~H., \& {Sigurdsson}, S. 2002, ApJ, 572, 556

\bibitem[{{Dufour} {et~al.}(2012){Dufour}, {Kilic}, {Fontaine}, {Bergeron},
  {Melis}, \& {Bochanski}}]{dufour_2012}
{Dufour}, P., {Kilic}, M., {Fontaine}, G., {et~al.} 2012, ApJ, 749, 6

\bibitem[{{Farihi}(2011)}]{farihi_mar2011}
{Farihi}, J. 2011, in American Institute of Physics Conference Series, Vol.
  1331, American Institute of Physics Conference Series, ed. S.~{Schuh},
  H.~{Drechsel}, \& U.~{Heber}, 193--210

\bibitem[{{Farihi} {et~al.}(2011){Farihi}, {Brinkworth}, {G{\"a}nsicke},
  {Marsh}, {Girven}, {Hoard}, {Klein}, \& {Koester}}]{farihi_feb2011}
{Farihi}, J., {Brinkworth}, C.~S., {G{\"a}nsicke}, B.~T., {et~al.} 2011, ApJl,
  728, L8

\bibitem[{{Farihi} {et~al.}(2012){Farihi}, {G{\"a}nsicke}, {Steele}, {Girven},
  {Burleigh}, {Breedt}, \& {Koester}}]{farihi_2012}
{Farihi}, J., {G{\"a}nsicke}, B.~T., {Steele}, P.~R., {et~al.} 2012, MNRAS,
  421, 1635

\bibitem[{{Farihi} {et~al.}(2010){Farihi}, {Jura}, {Lee}, \&
  {Zuckerman}}]{farihi_2010}
{Farihi}, J., {Jura}, M., {Lee}, J.-E., \& {Zuckerman}, B. 2010, ApJ, 714,
  1386

\bibitem[{{Farihi} {et~al.}(2009){Farihi}, {Jura}, \&
  {Zuckerman}}]{farihi_2009}
{Farihi}, J., {Jura}, M., \& {Zuckerman}, B. 2009, ApJ, 694, 805

\bibitem[{{Farihi} {et~al.}(2008){Farihi}, {Zuckerman}, \&
  {Becklin}}]{farihi_2008}
{Farihi}, J., {Zuckerman}, B., \& {Becklin}, E.~E. 2008, ApJ, 674, 431

\bibitem[{{Friedjung}(1985)}]{friedjung_1985}
{Friedjung}, M. 1985, \aap, 146, 366

\bibitem[{{G{\"a}nsicke} {et~al.}(2008){G{\"a}nsicke}, {Koester}, {Marsh},
  {Rebassa-Mansergas}, \& {Southworth}}]{gansicke_2008}
{G{\"a}nsicke}, B.~T., {Koester}, D., {Marsh}, T.~R., {Rebassa-Mansergas}, A.,
  \& {Southworth}, J. 2008, MNRAS, 391, L103

\bibitem[{{G{\"a}nsicke} {et~al.}(2007){G{\"a}nsicke}, {Marsh}, \&
  {Southworth}}]{gansicke_2007}
{G{\"a}nsicke}, B.~T., {Marsh}, T.~R., \& {Southworth}, J. 2007, MNRAS, 380,
  L35

\bibitem[{{G{\"a}nsicke} {et~al.}(2006){G{\"a}nsicke}, {Marsh}, {Southworth},
  \& {Rebassa-Mansergas}}]{gansicke_2006}
{G{\"a}nsicke}, B.~T., {Marsh}, T.~R., {Southworth}, J., \&
  {Rebassa-Mansergas}, A. 2006, Science, 314, 1908

\bibitem[{{Girven} {et~al.}(2012){Girven}, {Brinkworth}, {Farihi},
  {G{\"a}nsicke}, {Hoard}, {Marsh}, \& {Koester}}]{girven_2012}
{Girven}, J., {Brinkworth}, C.~S., {Farihi}, J., {et~al.} 2012, ApJ, 749, 154

\bibitem[{{Graham} {et~al.}(1990){Graham}, {Matthews}, {Neugebauer}, \&
  {Soifer}}]{graham_1990}
{Graham}, J.~R., {Matthews}, K., {Neugebauer}, G., \& {Soifer}, B.~T. 1990,
  ApJ, 357, 216

\bibitem[{{Guhathakurta} \& {Draine}(1989)}]{guhathakurta_1989}
{Guhathakurta}, P., \& {Draine}, B.~T. 1989, ApJ, 345, 230

\bibitem[{{Jura}(2003{\natexlab{a}})}]{jura_jan2003}
{Jura}, M. 2003{\natexlab{a}}, ApJ, 582, 1032

\bibitem[{{Jura}(2003{\natexlab{b}})}]{jura_feb2003}
---. 2003{\natexlab{b}}, ApJl, 584, L91

\bibitem[{{Jura}(2006)}]{jura_2006}
---. 2006, ApJ, 653, 613

\bibitem[{{Jura} {et~al.}(2007{\natexlab{a}}){Jura}, {Farihi}, \&
  {Zuckerman}}]{jura_jul2007}
{Jura}, M., {Farihi}, J., \& {Zuckerman}, B. 2007{\natexlab{a}}, ApJ, 663,
  1285

\bibitem[{{Jura} {et~al.}(2009){Jura}, {Farihi}, \& {Zuckerman}}]{jura_2009}
---. 2009, \aj, 137, 3191

\bibitem[{{Jura} {et~al.}(2007{\natexlab{b}}){Jura}, {Farihi}, {Zuckerman}, \&
  {Becklin}}]{jura_may2007}
{Jura}, M., {Farihi}, J., {Zuckerman}, B., \& {Becklin}, E.~E.
  2007{\natexlab{b}}, \aj, 133, 1927

\bibitem[{{Jura} \& {Xu}(2012)}]{jura_2012}
{Jura}, M., \& {Xu}, S. 2012, \aj, 143, 6

\bibitem[{{Jura} {et~al.}(2012){Jura}, {Xu}, {Klein}, {Koester}, \&
  {Zuckerman}}]{jura_may2012}
{Jura}, M., {Xu}, S., {Klein}, B., {Koester}, D., \& {Zuckerman}, B. 2012,
  ApJ, 750, 69

\bibitem[{{Kilic} {et~al.}(2012){Kilic}, {Patterson}, {Barber}, {Leggett}, \&
  {Dufour}}]{kilic_2012}
{Kilic}, M., {Patterson}, A.~J., {Barber}, S., {Leggett}, S.~K., \& {Dufour},
  P. 2012, MNRAS, 419, L59

\bibitem[{{Kilic} {et~al.}(2005){Kilic}, {von Hippel}, {Leggett}, \&
  {Winget}}]{kilic_2005}
{Kilic}, M., {von Hippel}, T., {Leggett}, S.~K., \& {Winget}, D.~E. 2005,
  ApJl, 632, L115

\bibitem[{{Kilic} {et~al.}(2006){Kilic}, {von Hippel}, {Leggett}, \&
  {Winget}}]{kilic_2006}
---. 2006, ApJ, 646, 474

\bibitem[{{Kimura} {et~al.}(2002){Kimura}, {Mann}, {Biesecker}, \&
  {Jessberger}}]{kimura_2002}
{Kimura}, H., {Mann}, I., {Biesecker}, D.~A., \& {Jessberger}, E.~K. 2002,
  Icarus, 159, 529

\bibitem[{{Klein} {et~al.}(2011){Klein}, {Jura}, {Koester}, \&
  {Zuckerman}}]{klein_2011}
{Klein}, B., {Jura}, M., {Koester}, D., \& {Zuckerman}, B. 2011, ApJ, 741, 64

\bibitem[{{Klein} {et~al.}(2010){Klein}, {Jura}, {Koester}, {Zuckerman}, \&
  {Melis}}]{klein_2010}
{Klein}, B., {Jura}, M., {Koester}, D., {Zuckerman}, B., \& {Melis}, C. 2010,
  ApJ, 709, 950

\bibitem[{{Koester} {et~al.}(2005){Koester}, {Rollenhagen}, {Napiwotzki},
  {Voss}, {Christlieb}, {Homeier}, \& {Reimers}}]{koester_2005}
{Koester}, D., {Rollenhagen}, K., {Napiwotzki}, R., {et~al.} 2005, \aap, 432,
  1025

\bibitem[{{Lodders}(2003)}]{lodders_2003}
{Lodders}, K. 2003, ApJ, 591, 1220

\bibitem[{{Melis} {et~al.}(2010){Melis}, {Jura}, {Albert}, {Klein}, \&
  {Zuckerman}}]{melis_2010}
{Melis}, C., {Jura}, M., {Albert}, L., {Klein}, B., \& {Zuckerman}, B. 2010,
  ApJ, 722, 1078

\bibitem[{{Metzger} {et~al.}(2012){Metzger}, {Rafikov}, \&
  {Bochkarev}}]{metzger_2012}
{Metzger}, B.~D., {Rafikov}, R.~R., \& {Bochkarev}, K.~V. 2012, MNRAS, 423,
  505

\bibitem[{{Podolak}(2010)}]{podolak_2010}
{Podolak}, M. 2010, in IAU Symposium, Vol. 263, IAU Symposium, ed. J.~A.
  {Fern{\'a}ndez}, D.~{Lazzaro}, D.~{Prialnik}, \& R.~{Schulz}, 19--28

\bibitem[{{Rafikov}(2011{\natexlab{a}})}]{rafikov_may2011}
{Rafikov}, R.~R. 2011{\natexlab{a}}, ApJl, 732, L3

\bibitem[{{Rafikov}(2011{\natexlab{b}})}]{rafikov_sep2011}
---. 2011{\natexlab{b}}, MNRAS, 416, L55

\bibitem[{Richter {et~al.}(2007)Richter, Janney, Mendybaev, Davis, \&
  Wadhwa}]{richter_2007}
Richter, F.~M., Janney, P.~E., Mendybaev, R.~A., Davis, A.~M., \& Wadhwa, M.
  2007, Geochimica et Cosmochimica Acta, 71, 5544

\bibitem[{{Zatisev} {et~al.}(2001){Zatisev}, {Shelkova}, {Litvina},
  {Sjakhpazov}, {Mogutnov}, \& {Syverud}}]{zatisev_2001}
{Zatisev}, A.~I., {Shelkova}, N.~E., {Litvina}, A.~D., {et~al.} 2001, High
  Temperature, 39, 388

\bibitem[{{Zuckerman} \& {Becklin}(1987)}]{zuckerman_1987}
{Zuckerman}, B., \& {Becklin}, E.~E. 1987, \nat, 330, 138

\bibitem[{{Zuckerman} {et~al.}(2007){Zuckerman}, {Koester}, {Melis}, {Hansen},
  \& {Jura}}]{zuckerman_2007}
{Zuckerman}, B., {Koester}, D., {Melis}, C., {Hansen}, B.~M., \& {Jura}, M.
  2007, ApJ, 671, 872

\bibitem[{{Zuckerman} {et~al.}(2003){Zuckerman}, {Koester}, {Reid}, \&
  {H{\"u}nsch}}]{zuckerman_2003}
{Zuckerman}, B., {Koester}, D., {Reid}, I.~N., \& {H{\"u}nsch}, M. 2003, ApJ,
  596, 477

\end{thebibliography}
\end{document}